\documentclass[aps,a4paper,
twocolumn,
rpsfig]{revtex4}

\usepackage{bm}
\usepackage{mathrsfs}
\usepackage{xcolor,color,graphicx,graphics}
\usepackage[all]{xy}
\usepackage{epsfig,subfigure}
\usepackage{latexsym,amssymb,amsmath,amsfonts} 
\usepackage[english]{babel} 
\usepackage[OT1]{fontenc}
\usepackage[latin1]{inputenc}
\usepackage{makeidx}
\usepackage{hyperref}
\usepackage{color,graphicx,graphics,wrapfig,epsf}

\begin{document}

\title{Scattering of GWs on wormholes: foreshadow and afterglow/echoes from
binary merges }
\author{A.A. Kirillov}
\email{ka98@mail.ru}
\affiliation{Bauman Moscow State Technical University, Moscow, 105005, Russian Federation}
\author{O.M. Lecian}
\email{orchideamaria.lecian@uniroma1.it, lecian@icra.it}
\affiliation{ICRA, ICRANeT, and Sapienza University of Rome, Via dei Marsi, 78- 00185
Rome, Italy}
\author{E.P. Savelova}
\email{savelova@bmstu.ru}
\affiliation{Bauman Moscow State Technical University, Moscow, 105005, Russian Federation}
\date{}

\begin{abstract}
We study specific collective features of the scattering of gravitational waves on wormholes and normal matter objects.
We derive and solve the GW energy transport equation and show that the scattered signal lies in the same
frequency spectrum bands as  the basic signal. The scattering forms  specific long living tails which always
accompany the basic signal and have a universal form.
The scattering on normal matter objects forms tails which have always the retarded character, while wormholes
lead to advanced tails as well. In addition, wormholes may produce considerably stronger effect when the total energy 
in tails  detected  on the Earth exceeds that in the incident direct wave  by the factor up to $10^{3}$.
In both cases the retarding tails have a long living character when the mean amplitude behaves with time
as  $h\sim 1/\sqrt{t+R/c}$. For a single GW event the echo tails give only a tiny contribution to the mean amplitude.
However such tails  accumulate with events and may be observed by their contribution to the noise
produced by the stochastic GW background.
\end{abstract}

\maketitle

\section{Introduction}

The gravitational-wave (GW) event GW150914 \cite{Abbot16} has opened a new
window in the Universe that gave rise to the born of the gravitational wave
astronomy. Further observations \cite{Abbot17} and new GW projects launched
(e.g., see  \cite{Kagra}) promise new and more detailed information
about the structure and different processes in our Universe. In particular,
several different groups have recently claimed tentative evidence for
repeating echoes in the LIGO/Virgo observations of binary black hole mergers
 \cite{echoes1,echoes2,echoes3,echoes4}. The last group 
 claims the
significance to reach $4.2\sigma $ at $f\simeq 72$ Hz. We leave aside
numerous possible mechanisms of origin of such echoes discussed in
literature (e.g., see,  \cite{Card16,Card19} and references in the above papers) and,  in the present paper,
explore properties of echoes which are generated by the scattering of the
basic GW signals on a distribution of relic wormholes and normal matter objects.
Namely, we consider collective effects of such a scattering which do not depend on the
exact specific structure of wormholes or other compact objects.

The simplest natural echo events may appear simply by lensing effects (e.g. see \cite{Lens01}).
Indeed, in considering the scattering of GWs on normal astrophysical objects
the most rough estimate gives for the cross-section the value of the order $%
\sigma _g\propto r_{g}^{2}$, where $r_{g}=2GM$ is the gravitational radius
of an object. This will cause the deflection angle $\theta \sim 2r_{g}/r$ ($r
$ is the distance from the ray to the object) and the time delay for
additional images of the source $\delta t/\Delta t\sim \theta ^{2}$ ($\Delta t$ is the
time passed from the emission to the detection). In general, for compact
objects all such values are extremely small. A considerable effects may
appear in two cases, when the distance from the ray to the object is
sufficiently small, or when the mass is sufficiently big. On cosmological
distances the time delay accumulates and becomes random. The accumulated
random delays affect also the primary ray, which sets
additional difficulties in observations of such effects. This means that in
the first place the GW astronomy allows to probe only very large
astrophysical objects (lensing on galaxies, clusters etc.).

 Despite to the fact that the creation of relic wormholes in the very early Universe
from quantum spacetime foam seem to be  a rather natural process (e.g., see  \cite{LQG1,LQG3}), 
wormholes are
still assumed to be exotic astrophysical objects. This happens, in the first
place, due to the absence of a direct observational evidence for wormholes.
Unlike normal objects wormholes possess the cross-section of the order $%
\sigma _w\propto b^{2}$, where $b$ is the throat radius, which can take
very big values $b\gg r_{g}$. In addition to lensing, it is possible to
observe a more detailed diffraction picture which appears due to the
scattering on a single wormhole \cite{Clement,KS18}. In this case the necessary condition
is to have a wormhole on the line of sight between the source of GW
radiation and the observer. That seems to be an extremely rare situation.
However, if we consider a distribution of relic wormholes, then the scattering of
GWs on wormholes should give a noticeable contribution to the scattered
signal. Indeed, in the models in which dark matter phenomenon appears due to
a distribution of relic wormholes \cite{KS07,KS11} the total scattered signal may
exceed the direct GW signal on the factor up to $e^{\tau _w}\sim 10^{3}$, where
$\tau _w$ is the mean optical depth for GWs. We recall that for the scattering
of GWs on normal astrophysical objects (e.g., stars) the optical depth is negligible $\tau _s
\sim \sigma _s n_s \ell\ll 1$,  where  $\sigma _s\sim <r_g^2>$ is the mean cross-section for a
typical star,  $n_s$ is the mean density of stars, and $\ell$ is the traversed path along the ray,
and the Universe is completely transparent (e.g., for $\ell =50Mpc$,  $\sigma _ s = \sigma _{\odot } $, 
and $ n_s =\rho _b /M_{\odot }$ we get
$\tau \sim  10^{-27 }$). In the presence of
wormholes however the Universe may be transparent only for some particular
directions/rays. 
In models where dark matter appears due to the presence of wormholes only  \cite{KS07,KS11}
the mean value of $\tau _w$ can be estimated by the
dark to luminous matter ratio (e.g., in low surface brightness galaxies the
mass to luminosity ratio $M/L$ may reach $10^{3}$ already at the edge of the
optical disk). We also point out that it is not difficult to suggest
parameters of a single wormhole that produce the scattered signal exceeding
or having the same order as the direct GW signal. The only problem here is
that the arrival times will be considerably separated (up to millions of years).

The estimate $\sigma _w\propto b^{2}$ follows from the geometrical optics 
which assumes wavelengths $\lambda \ll b$.
Therefore, for wavelengths obeying the above inequality
such an estimate is valid for all types of radiation scattered by wormholes (CMB, GWs,
Cosmic Rays, radiation emitted by galaxies, etc.) and results presented in the present paper
maybe equally applied to any type of radiation.
Effects of scattering of
radiation (different from GWs) on wormholes mix with analogous effects produced
 by normal matter \cite{KS18}. It is very
difficult to separate them. In this sense GWs give a unique tool to probe
the existence of wormholes in space, since they are weakly scattered by
ordinary matter and ordinary astrophysical objects. See however  \cite{KS20a} where possible effects of magnetic wormholes were discussed. 
In the present paper we
demonstrate that the scattering of the basic GW signal from a binary merge forms a specific
halo of secondary sources and at the Earth every GW impulse will have  heavy long-
living tails. In the case of wormholes there are both - advanced and retarded tails,
while the total energy flux
emitted by the halo and the energy contained in the tails may exceed that  in  the basic signal.
However this energy is widely distributed in time.

The two basic features
of such a scattering are that the scattered signal gets into the same frequency window as
the basic signal and that the mean GW amplitude $h$ ($h\sim \sqrt{I_{\omega} }$, where $ I_{\omega}\sim$
$\left\langle
\dot{h}^2\right\rangle =\frac{1}{T}\int_T (\dot{h}^2_\times +\dot{h}^2_+)dt$, $T=2\pi/\omega $, and $\dot{h}=\partial h/\partial t$)
in
the tail decays with time very slowly $\delta I_{\omega}\sim 1/(t+R/c)$,  
where $R$ is the distance to the GW source
and, therefore, $h(t)\sim 1/\sqrt{t+R/c}$.
The  amplitude ratio $\delta I/I^0$ has the order $\tau c\delta t/R\ll 1$, where $\tau $ is the optical depth 
and $\delta t$ is the duration of the basic signal and, therefore, the direct detection of such tails is not possible.
However, when taking into account all different binary merge events, tails accumulate which gives an additional multiplier $N\sim a\nu R/c\gg 1$, where
$\nu$ is the rate of events and the coefficient $a$ characterizes how long such tails live. Therefore, we should predict
essentially enhanced level of the stochastic GW background in the given
frequency band (roughly by the factor $\rho _{DM}/\rho _{b}$ as compared
with the measured/predicted rate of merges, where $\rho _{DM}$ and $\rho _{b}
$ are the dark matter and baryon densities).

The paper is organized as follows. In Sec. II we present an overview of existing results which concern the scattering on a single wormhole. In Sec. III we
describe  the topological structure of space in the presence of a distribution of relic wormholes. We also illustrate the origin of echoes for the exact model, i.e., a torus-like wormhole in the open Friedman model, and construct the map of geodesics for a spherical thin-shell wormhole which contains both entrances in the same space. In Sec. IV we derive the energy transport equation (ETE) for gravitational waves. The range of applicability of ETE is restricted by geometric optics (slow variation of the background metric as compared to the wavelengths) and it assumes random phases of GWs. In Sec. V we present solutions for ETE for a point-like GW source and in the approximation when the expansion of the Universe may be neglected. That corresponds to the local Universe with redshifts $z< 0.1$ We show that the scattering leads to the damping of the intensity along rays and forms a distributed halo of secondary sources which emit omnidirectional secondary flux. We also show how solutions generalize to the case of the expanding background (to an arbitrary redshifts). The secondary sources form tails for the direct basic signal. The structure of such tails is described in Sec. VI. In Sec. VII we discuss corrections due to possible particular motions of wormholes and the contribution from the Universe expansion.

\section{Scattering by a single wormhole:
 preliminary results
}
We point out that the problem of scattering of radiation by a single wormhole was first considered by Clement \cite{Clement}, 
see also references in \cite{KS18}.
In the case of GWs this problem involves several
difficulties. The most of existing models of wormholes involve exotic matter 
and the cross-section depends essentially on the structure of such a
wormhole (matter properties, structure of the throat, etc.). Some features
should not depend on the exact structure of wormholes and the matter
supporting the wormhole. Moreover, since GW sources (binary merges) are
rather rear events, it is an extremely rear chance to observe the scattering
on a single wormhole (it should have the position on the line of sight).
Part of this problem is treated in \cite{Kang2018,Kang:2019}.

It is possible to find a metric, for which the features of the wormhole do
not depend on the kind of matter supporting it. These kinds of metrics do
not even depend on the particular structure of the wormhole. After \cite{1},
\cite{2},  \cite{4}, \cite{5}, and, as summarized, for example, in
\cite{Kuhfittig:2009wm}, for a Schwarzschild-spherically-symmetric
Morris-Thorne wormhole, the metric chosen corresponds to the line element
\begin{equation}
ds^{2}=-e^{2\beta (r)}dt^{2}+e^{2\alpha (r)}dr^{2}+r^{2}(d^{2}\theta +\sin
\theta d\phi ^{2})  \label{MT}
\end{equation}%
where the shape function $b(r)$ is defined as
\begin{equation}
e^{2\alpha (r)}\equiv \frac{1}{1-\frac{b(r)}{r}}.
\end{equation}%
At the minimum radius $r_{0}$, the wormhole is defined by the function $%
\beta (r)$ assuming the value $\beta (r_{0})\equiv r_{0}$ (throat measure),
with the vertical asymptotics
\begin{equation}
\lim_{r\rightarrow {r_{0}}^{+}}\alpha (r)=+\infty
\end{equation}%
Further details can be found in \cite{Kim:2019xyf}, \cite{Kim:2019ojs} and
\cite{Kim:2017hem}.

The scattering of Gaussian gravitational wave packets by a single
Morris-Thorne wormhole has been studied in \cite{Ghersi:2019trn}, where the
transverse polarization component and the traceless polarization components
have been analyzed, both for the odd  and even potentials. 
The ECOs are
calculated by the geometrical-optic approximation and evaluated as produced
only in a narrow band, for which the detection would perturb the signal by
adding noise only in the asymptotical signal (outside the bandwidth
evaluated). As a
result, only a very narrow model-dependent band peak has been demonstrated
to be observable.

For a single Kerr-like wormhole, whose perturbations are ruled by a
symmetric potential, the echoes of a scattered signal have been analyzed
with similar results in \cite{Bueno:2017hyj}; rotation has been demonstrated
to modify the emission forms.
The observational differences between the scattering of gravitational
radiation on a single wormhole and that on ultra-compact stars was studied by the
differences of the quasi-normal modes \cite{Volkel:2018hwb} by means of the
inverse-spectrum method for the detection of gravitational waves, for which
non-negligible differences arise in the non-negligible experimental errors
for next generation gravitational wave detectors in the reconstruction of
the quasi-normal spectrum.

For single ultra-static wormholes, non-Schwarzschild symmetric wormhole
metrics have been demonstrated to reduce the necessity for the explanation
of the wormhole properties by exotic matter, and the guidelines to eliminate
such dependence on the specific matter content rather than on the solution
to the Einstein field equations for the Schwarzschild-symmetric case have
been outlined \cite{Visser:1989kh}.

A lensing for the shadow of Misner-Thorne wormholes is proposed in \cite%
{lens1}, in the case of spherical symmetry. The edge of the lensing allows
one to distinguish the wormhole form other celestial bodied. The analysis of
the lensing is also suited for the analysis of the parameters qualifying the
solutions to the EFE's.

The strong gravitational lensing for static wormhole geometries is analyzed
in \cite{lens2}. In particular, the parameters available in the lensing
effect provide one with the possibility to determine different wormhole
geometries and different celestial-bodies geometries.

In \cite{lens3}, the differences in the observational evidence for a
Schwarzschild wormhole, an Ellis-Bronnikov wormhole, a Schwarzschild
blackhole and a Kerr blackhole are studied. The estimations of the parameters
characterizing the solutions for the EFE's of an Ellis-Bronnikov wormhole
are constrained by experimental accuracy and the criteria to keep it
distinguished from a Schwarzshild blackhole are determined in the
gravitational strong-lensing regime after the study of the quasi-normal
modes. 

In \cite{lens4}, the gravitational lensing for an Ellis wormhole is
calculated. The case of a collection of wormhole is also faced. In
particular, the radius of the throat of the wormhole is taken as a
determining factor for the determination of the number density of the
collection of wormholes for identifying it form other compact objects in the
geometrical-optics approximation. The observational methods of femto-lensing
analysis \cite{lens41}-\cite{lens41b}, micro-lensing and or the astrometric-
image-centroid- displacements analysis \cite{lens42} are compared and found
not suited for the analysis of such macroscopic celestial configurations, as
they do not resolve the modulation of the light curve or the displacements
in the time domain.

In \cite{lens5}, the geodesics and the deflection angle for a charged
Loretzian wormhole \cite{lens51} are calculated.
In \cite{lens6}, the observables for the strong gravitational lensing of
wormholes are determined. The Ellis-Bronnikov solutions is analyzed.
Different locations for the locations are proposed in the vicinity of the
wormhole, which are however not compatible with Earth-based experiments
and/or with satellite-based experiments.

Thus all the previous results agree that scattering by a single wormhole 
either produces too small effect to be observed by existing experiments, or requires 
to realize an extremely specific configuration involving parameters and positions of a GW source and a wormhole.

\section{Topological structure of space with a distribution/gas of relic wormholes}

In general relativity topological structure of space is determined by onset,
as additional initial conditions \cite{Geroch} (see also recent discussion
in  \cite{KS20}). The most natural initial conditions correspond to a
fractal topological structure of space, i.e., to a fractal distribution of
wormholes. Indeed, lattice quantum gravity models predict fractal properties
of space at sub-planckian scales (e.g., see  \cite{LQG1,LQG3}), while the inflationary phase in
the past should enormously stretch all scales and temper such a structure
as initial conditions.

We point out that the space which contains a distribution of wormholes
cannot be covered by a single coordinate map. In astrophysical applications
however it is convenient to describe wormholes in terms of the single
coordinate map which is commonly used to describe points in the homogeneous
and isotropic Universe (e.g., the red shift and two angles on the Sky, or any other
coordinates of the flat Friedman model). This is achieved by making cuts
along the minimal sections of wormhole throats. As the result we get the
space manifold with a set of couples of boundaries at which we should specify
specific boundary conditions. Boundary conditions follow simply from the
fact that all physical fields are continuous at
throats. On sky such boundaries will be seen as couples of specific/exotic
astrophysical objects which in general have a rather complex form (couples
of two-dimensional surfaces $S_{A}$, where the index $A=1,..N$ numerates
wormhole throats). Those surfaces/objects may move, rotate, possess equal
masses and magnetic poles. The boundary conditions are induced by the
cutting procedure. For example, if we consider a couple of such surfaces $%
S_{\pm }$ which correspond to the same wormhole, then the internal region of
space restricted by $S_{+}$ (or, in general, some part of it) admits a
one-to-one map on a portion of the outer region for the conjugated surface $%
S_{+}$. This defines the boundary conditions in a unique way.

In general case throat sections (their space-like part) have the form of a sphere
with $n$ handles $S_{n}^{2}$. As it was discussed in  \cite{KS16} the
simplest wormhole has the section which has the form of a torus. We recall
that such wormholes do not require the presence of exotic forms of matter
and they do evolve. In open Friedman models their rate of evolution exactly
coincides with the common cosmological expansion and they are stable. In the
case of the flat space the rate of their evolution is still not investigated
properly. Some encouraging results were obtained in the limit when one
radius of a torus-like throat tends to infinity and the geometry acquires
cylindrical configuration. In particular, in a series of papers 
\cite{Bron6,Bron9,Bron7,Bron8} static and stationary cylindrical wormhole
solutions were found and it was demonstrated that asymptotically flat
wormhole configurations do not require exotic matter violating the weak
energy condition. We point out that if such a torus-like wormhole rotates in
space, this surely should prevent it from very fast collapse. The thorough
investigation of such objects represents too complex problem which still
waits for its investigation.

In the present paper we, for the sake of simplicity, assume spherical
throats of wormholes. Some words to approve the use of such an approximation
worth adding. Indeed, spherical wormholes collapse very quickly and,
therefore, they cannot be distinguished from ordinary black holes. Stable
traversable spherical wormholes may exist only in the presence of exotic
forms of matter and in modified theories, where the role of the exotic
matter is played by an appropriate modification
\cite{WMod:HPS97,WMod:HLS13,WMod:MSV16}. So far, there is no any
rigorous experimental evidence for the existence of exotic forms of matter,
or for the presence of any modification of general relativity (we leave
aside possible quantum corrections which work only in the quasi-classical
region, e.g., at Planck scales). This means that the approximation of
spherical throats should be very rough. Nevertheless, in considering
collective effects of GW scattering on a distribution of wormholes, such an
approximation works rather well. Owing to the fact that more general
wormhole throats have random orientations in space and assuming that they
have an isotropic distribution, the final result always contains the
averaging over orientations. When we perform such an averaging, then
every throat restores the spherical symmetry. The spherical wormhole, in
turn, is much more simple object to work with and it admits a much more
simple consideration.

In conclusion of this section we consider the exact illustrative model of
the scattering on a single torus-like wormhole in the open model (see the exact
description of the appropriate metric in \cite{KS16,KS20}) and construct the map
of geodesic lines at boundaries for a spherical wormhole.

\subsection{Echoes from a single wormhole}

Consider first the open Freedman model. Then the space-like part represents
the standard Lobachevsky space. The simplest wormhole which connects two
Lobachevsky spaces is obtained by a factorization over a discrete subgroup
of the group of motions. The discrete subgroup $G$ is determined by a
couple of generators $T_{a}(l_{1})$ and $T_{b}(l_{2})$ which describe two
shifts of the space in orthogonal directions ($l_{1}$ and $l_{2}$ denote two
orthogonal geodesics) on distances $r_{1}=a$ and $r_{2}=b$. Any element $%
g_{A}\in G$ may be constructed merely as $g_{A}=T_{a}^{m}T_{b}^{n}$ with $m,n=0,\pm 1,\pm
2,...$. We point out that the shifts $T_{a}(l_{1})$ and $T_{b}(l_{2})$ do not commute and elements
$g\in G$ are classified by a more complicated way. This however is not important for subsequent consideration.

The minimal section of the throat has the form of a torus with two
radii $r_{1}$ and $r_{2}$. As it was demonstrated in \cite{KS20} it is
possible to introduce such coordinates on the Lobachevsky space $\left( \chi
_{1},\chi _{2},\chi _{3}\right) $ that the shift is merely $g_{A}\left(
\chi _{1},\chi _{2},\chi _{3}\right) =\left( \chi _{1}+2\pi m,\chi _{2}+2\pi
n,\chi _{3}\right) $, while perturbations of the metric $\delta g_{\alpha
\beta }(x)$ obey the "periodic" property
\begin{equation}
\delta g_{\alpha \beta }\left( \chi _{1},\chi _{2},\chi _{3}\right) =\delta
g_{\alpha \beta }\left( \chi _{1}+2\pi m,\chi _{2}+2\pi n,\chi _{3}\right).
\end{equation}

Consider now the retarded Green function $G_{F}(x,t;x^{\prime },t^{\prime })$
for the open Friedman model which describes the propagation of a single GW
impulse. The above periodicity means that in the presence of the wormhole
any source in the Friedman space also obeys the periodic conditions and,
therefore, in terms of the unrestricted Friedman space it induces multiple
additional images%
\begin{equation}
\frac{1}{\sqrt{\gamma }}\delta _{w}\left( x-x^{\prime }\right) =\frac{1}{%
\sqrt{\gamma }}\delta _{F}\left( x-x^{\prime }\right) +
\sum_{A\not=0}\frac{%
1}{\sqrt{\gamma }}\delta _{F}\left( x-g_{A}(x')\right).
\label{DMS}
\end{equation}%
Then in the presence of the wormhole the true Green function has the obvious
structure%
\begin{equation}
G_{w}(x,t;x^{\prime },t^{\prime })=G_{F}(x,t;x^{\prime },t^{\prime
})+\sum_{A\not=0}G_{F}(x,t;g_{A}(x^{\prime }),t^{\prime }).  \label{GFE}
\end{equation}%
The same structure acquires any signal emitted by binary merges. Here the
first term corresponds to the direct GW signal, while the sum corresponds to
echoes. The amplitudes of echoes depend on the position of the source (with
respect to the wormhole center) and the parameters of the wormhole (radii $a$
and $b$). We point out that the general structure of echoes (\ref{GFE})
appears also for spherically symmetric wormholes in the asymptotically flat
space, e.g., see  \cite{Ghersi:2019trn,Bueno:2017hyj}.
Therefore, it is not a specific feature of the model discussed. When we have
a distribution of such wormholes the general structure remains the same,
while the GW signal becomes much more complicated. Moreover, part of echoes
overrun the basic signal. In general amplitudes of echoes are extremely small
and phases become random.

It was shown in  \cite{KS07,KS11} that additional images of an actual
source (\ref{DMS}) may play the role of dark matter. In this case the
distribution of DM in the Universe is determined by the distribution of
wormholes, while the observed in galaxies rigid relation between visible and
dark matter components
acquires the most natural explanation. This means also
that the multiple echoes have the same origin and the intensity of echoes
relates somehow to the distribution of dark matter. This feature will be
used for estimates of the intensity of the stochastic GW background.

\subsection{Geodesic map for spherical wormholes}

As it was pointed out the space with wormholes cannot be covered by a single
coordinate atlas. For practical aims it is convenient to have a single atlas
(e.g., a part of the Friedman space). In this case we cut the wormhole
throat by the minimal section of the throat. Then the wormhole represents a
couple of surfaces $\Sigma _{\pm }=S_{\pm }\times R$ (which are the direct
product of a sphere and the time axis) whose internal regions are removed,
while surfaces of the spheres are glued. From the topological standpoint
this corresponds to the standard Minkowski space with boundaries. In general
case for a remote observer such boundaries move in space and are glued by the Poincare motion $%
x_{+}^{\alpha }-X_{+}^{\alpha }=\Lambda _{\beta }^{\alpha }\left(
x_{-}^{\beta }-X_{-}^{\beta }\right) $. The gluing means that when the ray $%
x(\ell )$ reaches one such a boundary $\Sigma _{-}$ at some finite value $%
\ell _{0}$, i.e., $x_{-}^{\alpha }(\ell _{0})\in \Sigma _{-}$ and particular
values of $k_{\mu }^{-}(\ell _{0})$, it's continuation comes from the other
boundary $\Sigma _{+}$ with  new initial data $x_{+}^{\alpha }(\ell
_{0})\in \Sigma _{+}$ and $k_{\mu }^{+}(\ell _{0})$ related by the Poincare
map $x_{+}^{\alpha }-X_{+}^{\alpha }=\Lambda _{\beta }^{\alpha }\left(
x_{-}^{\beta }-X_{-}^{\beta }\right) $ and $k_{+}^{\alpha }=\Lambda _{\beta
}^{\alpha }k_{-}^{\beta }$. In general, there is some back reaction (some
change in parameters $X_{+}^{\alpha }$, $X_{-}^{\beta }$, and $\Lambda
_{\beta }^{\alpha }$) as described in  \cite{KS11}. In the present paper
we neglect the back reaction. In the reference frames in which the throat
entrances are at rest the map corresponds simply to the inversion of the
spheres. This can
be easily seen for the simplest Ellis-Bronnikov metric (EB-wormhole)
$ds^2 =dt^2-f^2(r)dl^2$,
where $f(r)=1+b^2/r^2$ and $dl^2=dr^2+r^2(\sin ^2\theta d\phi ^2+d\theta ^2)$
is the standard line element of the flat space. The inversion map $r'=b^2/r$ simply interchanges
the inner and outer  regions of the sphere $r=b$ which is the minimal throat section
of the EB-wormhole. Close to the minimal section $r=b$ almost any spherical
metric with a smooth source can be reduced to this case.

First, for the sake of simplicity we assume that the space is flat (this is
not an approximation since in the neighborhood of any point of the surfaces $%
\Sigma _{\pm }$ the metric can be taken as a flat Minkowski metric, the only
exclusion case is the so-called thin-shell wormholes). Second, we assume $%
X_{\pm }^{\alpha }=(t,\mathbf{R}_{\pm })$ (the shift of time is absent), and
velocities $\frac{d}{dt}\mathbf{R}_{\pm }=\mathbf{V}_{\pm }$ $\ll c$
(velocities of centers of spheres $S_{\pm }$). Then $\Lambda _{\beta
}^{\alpha }$ is a composition of a space rotation $U_{\beta }^{\alpha }$ and
Lorentz boost. 

Consider the incident wave $(\omega _{in},\mathbf{k}_{in})$.
Let it falls on the throat $S_{-}$ at the point $\mathbf{x}_{-}^{\prime }=b%
\mathbf{n}_{-}^{\prime }+\mathbf{R}_{-}$ (where $b\mathbf{n}_{-}^{\prime }=%
\mathbf{\xi }_{-}^{\prime }=\mathbf{x}_{-}^{\prime }-\mathbf{R}_{-}$). Since
the throat $S_{-}$ moves in space with the velocity $\mathbf{V}_{-}$ the
frequency and the wave number in the coordinate system in which the throat $%
S_{-}$ is at rest are (we assume $ V_{\pm }/c\ll 1$)
\begin{equation}
\omega _{-}^{\prime }\simeq \omega _{in}-\left( \mathbf{V}_{-}\mathbf{k}%
_{in}\right) ,\ \ \mathbf{k}_{-}^{\prime }\simeq \mathbf{k}_{in}\mathbf{-}%
\frac{\omega _{in}}{c}\frac{\mathbf{V}_{-}}{c}.  \label{S1}
\end{equation}%
This wave (ray) is absorbed by the throat $S_{-}$ and re-radiates from $%
S_{+} $ at the point $\mathbf{x}_{+}^{\prime }=b\mathbf{n}_{+}^{\prime }+%
\mathbf{R}_{+}$ which relates to $\mathbf{x}_{-}^{\prime }$ by the relation
(rotation)
\begin{equation*}
n_{+}^{l\prime }=U_{m}^{l}n_{-}^{m\prime }.
\end{equation*}%
In the reference system in which $S_{+}$ is at rest the frequency and the
wave number of the outgoing (re-radiated) wave are%
\begin{equation*}
\omega _{+}^{\prime }=\omega _{-}^{\prime },\ \ k_{+}^{l\prime
}=U_{m}^{l}\left( k_{-}^{m\prime }-2n_{-}^{m\prime }\left( \mathbf{n}%
_{-}^{\prime }\mathbf{k}_{-}^{\prime }\right) \right) .
\end{equation*}%
Thus in the initial coordinate system we find the re-radiated values $%
(\omega _{out},\mathbf{k}_{out})$
\begin{equation}
\omega =\omega _{out}\simeq \omega _{+}^{\prime }-\left( \mathbf{V}_{+}%
\mathbf{k}_{+}^{\prime }\right) ,\ \ \mathbf{k}=\mathbf{k}_{out}\simeq
\mathbf{k}_{+}^{\prime }\mathbf{-}\frac{\omega _{+}^{\prime }}{c}\frac{%
\mathbf{V}_{+}}{c}.  \label{S4}
\end{equation}

\section{Energy transport equation}

The incoherent nature of the scattered signal shows that the exact form of
the wavefront is not important for observations. It may be important only
for observing the scattering on a particular single wormhole. If we a so
happy to make all the necessary conditions meet, then we will need the
detailed structure of the scattered signal. However the situation is such
that in the nearest future we may hope to detect only collective effects of
their scattering. To this end the most convenient way to use the equations
for the transport of energy. Such an equation comes out from the standard
kinetic equation for the number of gravitons $N(k,r)$ in the phase space $%
\Gamma =(k,r)$. Indeed, in the case when phases of GWs are random the energy
of the GWs can be written as follows
\begin{equation}
W=\sum_{j}\int W_{k,j}d^{3}kd^{3}r=\sum_{j}\int \hbar \omega
N_{k,j}d^{3}kd^{3}r.  \label{Energy}
\end{equation}%
In the above expression the index $j$ stands for polarizations. In what follows we will omit
the index $j$. In equation (\ref{Energy}) $W_{k}$ is the spectral energy
density (energy in a unit volume of space and a unit volume of wave
numbers). In the isotropic case (more generally, when the number of
gravitons depends only on the frequency $N_{\omega }$) the spectral energy
is described by $W_{k}d^{3}k$ $=W_{k}k^{2}dkd\Omega $ $=W_{k}k^{2}\left\vert \frac{%
dk}{d\omega }\right\vert d\omega d\Omega $ $=W_{\omega }d\omega d\Omega $. When
considering radiation it is commonly used the spectral intensity of GW
radiation (Poynting vector or the energy flux)%
\begin{equation}
\mathbf{S}_{k}=W_{k}\frac{d\omega }{d\mathbf{k}},\ \ I_{k}=\left\vert
\mathbf{S}_{k}\right\vert ,
\end{equation}%
where $\frac{d\omega }{d\mathbf{k}}=\mathbf{V}_{g}$ is the group velocity.
In general relativity $\left\vert \mathbf{V}_{g}\right\vert =c$, while in
different modified theories its value may change. We see that intensity of
waves relates to $W_{k}$, $W_{\omega }$ simply as $I_{k}=cW_{k}$.

Consider the number of gravitons/photons ($W_{k}=\hbar \omega (k,r)N_{k}$)%
\begin{equation}
N_{k,r}=\sum_{a}\frac{1}{\sqrt{\gamma }}\delta \left( \mathbf{r}-\mathbf{r}%
_{a}(t)\right) \delta \left( \mathbf{k}-\mathbf{k}_{a}(t)\right) .
\end{equation}%
The equation for $N_{k}$ may be obtained from the geometric optics for
simplest case of a set of gravitons. One may consider gravitons as massless
(spin-2) particles with the momenta $p_{\mu }=\hbar k_{\mu }$ and the
dispersion relation $k_{\mu }k_{\nu }g^{\mu \nu }=0$ which defines the
energy $\omega (k,r,t)$. Here $\gamma _{ij}$ is the space metric, $\left(
\mathbf{r}(t),\mathbf{k}(t)\right) =$ $\Gamma _{k}(t)$ corresponds to a
particular isotropic geodesic line and $\omega (k,r,t)$ obeys to
\begin{equation}
g^{\alpha \beta }\frac{\partial \psi }{\partial x^{\alpha }}\frac{\partial
\psi }{\partial x^{\beta }}=g^{\alpha \beta }k_{\alpha }k_{\beta }=0.
\end{equation}%
Then the kinetic equation has the form (e.g., see  \cite{KS11})
\begin{equation}
\begin{array}{ll}
\frac{DN_{k}}{dt}
=\widetilde{\alpha }%
_{k}+\int \Sigma 
( \Gamma ,\Gamma _{w},\Gamma ^{\prime },\Gamma
_{w}^{\prime }
) 
N\left( \Gamma ^{\prime }\right) N_{w}\left( \Gamma
_{w}^{\prime }\right) d\Gamma ^{\prime }d\Gamma _{w}^{\prime }d\Gamma _{w},
\end{array}
\end{equation}%
where we use the definition
\begin{equation}
\begin{array}{ll}
\frac{DN_{k}}{dt}=\frac{1}{\sqrt{%
\gamma }}\frac{\partial \sqrt{\gamma }N_{k}}{\partial t}+\frac{1}{\sqrt{%
\gamma }}\frac{\partial }{\partial \mathbf{r}}\left( \frac{d\mathbf{r}}{dt}%
\sqrt{\gamma }N_{k}\right) +
\frac{1}{\sqrt{\gamma }}
\frac{\partial }{%
\partial \mathbf{k}}\left( \frac{d\mathbf{k}}{dt}
\sqrt{\gamma }
N_{k}\right),
\end{array}
\end{equation}%
$\widetilde{\alpha }_{k}$ is the emission of particles/gravitons in the
unite time and unite volume, $\Sigma $ is the scattering matrix
\begin{equation}
\Sigma \left( \Gamma ,\Gamma _{w},\Gamma ^{\prime },\Gamma _{w}^{\prime
}\right) =c\sigma \left( \Gamma ,\Gamma _{w},\Gamma ^{\prime },\Gamma
_{w}^{\prime }\right) .  \label{sigm}
\end{equation}%
In the above expression $\sigma \left( \Gamma ,\Gamma _{w},\Gamma ^{\prime
},\Gamma _{w}^{\prime }\right) $ is the cross-section of the scattering on a
wormhole $\Gamma ,\Gamma _{w}\rightarrow \Gamma ^{\prime },\Gamma ^{\prime
}{}_{w}$, (here $\Gamma $ and $\Gamma _{w}$ are parameters before the
scattering (incident) and $\Gamma ^{\prime }$ and $\Gamma _{w}^{\prime }$
are parameters after the scattering) which are determined by (\ref{S1})-(\ref%
{S4}), $N_{w}\left( \Gamma _{w}\right) $ is the number of wormholes in the
configuration space $\Gamma _{w}$, and $\Gamma _{w}=\left( R_{\pm },V_{\pm
},b,U,...\right) $ are all the parameters of the wormholes. In what follows
we will assume the case when wormhole are infinitely heavy objects (neglect
the back reaction), then we get
\begin{equation*}
\sigma \left( \Gamma ,\Gamma _{w},\Gamma ^{\prime },\Gamma _{w}^{\prime
}\right) =\widetilde{\sigma }\left( \Gamma ,\Gamma ^{\prime },\Gamma
_{w}\right) \delta \left( \Gamma _{w}-\Gamma _{w}^{\prime }\right) .
\end{equation*}%
In the general case it is given by $\sigma =\sigma _- +\sigma _+$ (e.g., see  \cite{KS11,KSZ08b})%
 \begin{equation}
 \begin{array}{cc}
\sigma _{-}\left( \Gamma ,\Gamma _{w},\Gamma ^{\prime },\Gamma _{w}^{\prime
}\right) 
=
\delta \left( \xi _{+}-b\right) \delta \left( \Gamma ^{\prime
}-\Gamma _{+}\right) \delta \left( \Gamma _{w}-\Gamma _{w}^{\prime }\right) 
-
\\
-
\delta \left( \xi _{-}-b\right) \delta \left( \Gamma ^{\prime }-\Gamma
\right) \delta \left( \Gamma _{w}-\Gamma _{w}^{\prime }\right)  ,
\end{array}
\label{sec}
\end{equation}
where $\Gamma ^{\prime }=\Gamma _{in}=(x_{in},\mathbf{k}_{in})$ and $\Gamma
=\Gamma _{out}=(x_{out},\mathbf{k}_{out})=(r,\mathbf{k})$. The sign $\sigma
_{-}$ means here that GW ray falls on $S_{-}$. The second term in (\ref{sec}%
) corresponds to the absorption on $S_{-}$ and the first term corresponds to
the reradiating of the absorbed signal from $S_{+}$. Analogous term $\sigma
_{+}$ corresponds to the wave which falls on $S_{+}$ and re-radiates from $%
S_{-}$. In the final expression (\ref{TEQ}) it gives only the factor $2$
(due to the symmetry between entrances).

The equation for the energy transport is found to be%
\begin{equation}\begin{array}{ll}
\frac{DW_{k}}{dt}-\left( \frac{1}{%
\omega }\frac{d\omega }{dt}\right) W_{k}=\alpha _{k}-\int \mu \left( \Gamma
_{k},\Gamma _{k}^{\prime }\right) W\left( \Gamma _{k}^{\prime }\right)
d\Gamma _{k}^{\prime } , 
\end{array}
\label{FEE}
\end{equation}%
where
\begin{equation}\label{MU1}
\mu \left( \Gamma _{k},\Gamma _{k}^{\prime }\right) =-\int \frac{\omega }{%
\omega ^{\prime }}\Sigma \left( \Gamma ,\Gamma ^{\prime },\Gamma _{w}\right)
N_{w}\left( \Gamma _{w}\right) d\Gamma _{w},
\end{equation}%
$\Sigma $ is defined by (\ref{sigm}), and the energy density $W_{k}$ for a
set of gravitons is
\begin{equation*}
W_{k}=\sum_{a}\frac{\hbar \omega (k_{a},r_{a},t)}{\sqrt{\gamma }}\delta
\left( \mathbf{r}-\mathbf{r}_{a}(t)\right) \delta \left( \mathbf{k}-\mathbf{k%
}_{a}(t)\right) .
\end{equation*}%
In equation (\ref{FEE}) the term $\alpha _{k}=\hbar \omega \widetilde{\alpha
}_{k}$ describes spontaneous radiation and $\mu $ describes
adsorption/re-radiation (induced radiation) of the GW radiation in a unit
volume of the medium.

We point out that equation  (\ref{FEE})  works also in the case when the medium
consists of normal matter objects (stars or black holes, gas, etc.). In such a  case however the scattering matrix
$\Sigma $ should be determined separately and in (\ref{MU1}) $N_w\rightarrow N_s$.
There exist however a phenomenological way to get estimates for the GW scattering on stars or black holes.
It corresponds to the limit when the separation of wormhole entrances vanishes $R_+=R_-$ and the throat radius
becomes the gravitational radius of the object $b=r_g$. Then the scattering laws (\ref{S1}) - (\ref{S4}) correspond simply to the
reflection of rays from the gravitational radius of the object. This case does not require a separate consideration,
since it can be modeled by a specific form
of the wormhole distribution $N_w(\Gamma _{w})=\tilde{ N}_s(\Gamma _-)\delta (\mathbf{R}_+ -\mathbf{R}_-)$, where
$\Gamma _-$ corresponds to the set of parameters of a single wormhole entrance.

The second term in l.h.s. of (\ref{FEE}) describes the change of the
energy flux due to the non-stationarity of the background. In particular, in
an expanding Universe the frequency changes according to the cosmological
shift only $\frac{1}{\omega }\frac{d\omega }{dt}=-H$ (here the dependence on
time goes through $\lambda (t)\sim r=a(t)x$) and the change of the volume
element is $\frac{1}{\sqrt{\gamma }}\frac{d\sqrt{\gamma }}{dt}=3H$, where $H$
is the Hubble constant ($\sqrt{\gamma }=a^{3}(t)$). If we neglect the
effects of the expansion (the red shift of the frequency), then (\ref{FEE})
reduces merely to
\begin{equation}
\frac{\partial W_{k}}{\partial t}+V_{g}\frac{dW_{k}}{d\ell }=\alpha
_{k}-\int \mu W_{k}^{\prime }d\Gamma _{k}^{\prime } ,  \label{FEE2}
\end{equation}%
where we used the relation $\frac{d\ell }{dt}=V_{g}=\left\vert \frac{d\omega
}{d\mathbf{k}}\right\vert $, $\ell $ is the natural parameter along the ray
(the length), and $\frac{d}{d\ell }=\frac{d\mathbf{r}}{d\ell }\frac{\partial
}{\partial \mathbf{r}}+\frac{d\mathbf{k}}{d\ell }\frac{\partial }{\partial
\mathbf{k}}$.

 The aim of this section was to derive the energy transport equation which is given by (\ref{FEE}). 
It is important that the structure of the scattering term (\ref{MU1}) which is determined by the scattering matrix (\ref{sec}) 
provides the rigorous conservation law for the number of absorbed and emitted gravitons. 
The energy of gravitons however depends on the redshift and the total energy does not conserve. 
It conserves only when we apply those equation to the local Universe with redshifts $z< 0.1$ (distances up to a billion of light years). In this case
 we can neglect the expansion  and use the more simple equation (\ref{FEE2}).

\section{Scattering of gravitational waves}
\subsection{Direct signal}

Let the spacetime be flat $\omega =ck$ and let us take the source in the
form
\begin{equation}
\alpha _{k}=w_{0}\left( k\right) \delta \left( \mathbf{r}-\mathbf{r}^{\prime
\prime }\right) \delta \left( t-t^{\prime \prime }\right)  . \label{Sc}
\end{equation}%
Then at the moment $t^{\prime \prime }$ we have radiation with the spectrum
$w_{0}(k) =L \delta \left( \omega -\omega _{0}\right) $ from the point $x^{\prime \prime }$. We point out that binary
merges form the source in the form
\begin{equation}
\alpha _{k}=\frac{dw\left( \omega ,t\right) }{dt}\delta \left( \mathbf{r}-%
\mathbf{r}^{\prime \prime }\right) ,\ \ \ w\left( \omega ,t\right)
=L(t)\delta \left( \omega -\omega _{0}(t)\right),   \label{Sc2}
\end{equation}%
where the amplitude $L(t)=Q(t)^{2}$ and $\omega _{0}(t)$ correspond to the
chirp signal \cite{chirp2}. The real signal can be simply taken as the sum of type (%
\ref{Sc}) signals.

In the leading order, without taking into account the scattering on
wormholes, we find the solution of (\ref{FEE2}) in the form  (see Appendix \ref{freem})
\begin{equation}
\begin{array}{ll}
W_{k}^{0}\left( x,t\right) =\frac{w_0\left( \omega \right) }{cR^2}\delta \left(
t^{\prime \prime }-t+\frac{R}{c}\right) \delta \left( \cos \theta -\cos \theta ^{\prime
}\right) \delta \left( \phi -\phi ^{\prime }\right),
\end{array}
\end{equation}%
where $R=\left\vert
\mathbf{r}-\mathbf{r}^{\prime \prime }\right\vert$ and  $\theta ^{\prime }$, $\phi ^{\prime }$ relate to the direction of the
velocity $\mathbf{V}_{g}$. The chirp signal (\ref{Sc2}) forms the energy
density as follows%
\begin{equation}\label{signal1}
W^0_k\left( r,t\right) =\frac{\left[ w\left( \omega ,t\right) \right] _{ret}}{%
cR^2
}\delta
\left( \cos \theta -\cos \theta ^{\prime }\right) \delta \left( \phi -\phi
^{\prime }\right)
\end{equation}%
where we denote $\left[ f(t)\right] _{ret}=f\left( t-\frac{\left\vert
\mathbf{r}-\mathbf{r}^{\prime \prime }\right\vert }{c}\right) $.

\subsection{Damping and echoes: the halo of secondary sources }

The additional signal comes from the additional sources (the so-called a
distributed hallo)%
\begin{equation}
\delta \alpha _{k}=\int \frac{ \omega }{ \omega ^{\prime }}\Sigma \left( \Gamma ,\Gamma
_{w},\Gamma ^{\prime },\Gamma _{w}^{\prime }\right) W_k^{0 \prime}
 N_{w}^{\prime } d\Gamma
^{\prime }d\Gamma _{w}^{\prime }d\Gamma
_{w}.
\end{equation}%
If the distribution of wormholes in terms of the matrix $U$ is isotropic,
then re-radiation will come out in an isotropic way (omnidirectional flux).
Then we should follow only the frequency shift which is given by (\ref{S1}), (\ref{S4})
\begin{equation}
\omega _{out}\simeq \omega _{in}\left( 1-\frac{1}{c}\left( %
\mathbf{V}_{-}\mathbf{m}_{in}\right) -\frac{1}{c}\left( \mathbf{V}_{+}\mathbf{m}%
_{out}\right) \right) .\
\end{equation}%
We point out that $in$ and $out$ states are symmetric. Here the unit vector $%
\mathbf{m}_{in}=\left( \mathbf{R}_{-}-\mathbf{x}^{\prime \prime }\right)
/\left\vert \mathbf{R}_{-}-\mathbf{x}^{\prime \prime }\right\vert $ and $%
\mathbf{m}_{out}=\left( \mathbf{x}-\mathbf{R}_{+}\right) /\left\vert
\mathbf{x-R}_{+}\right\vert $ points to the observer.

Let us evaluate the term which corresponds to the absorption of radiation.
We point out that the absorption term in (\ref{sec}) defines simply damping
of the intensity and has the structure in (\ref{FEE2})
\begin{equation}
\int \mu W_{k}^{\prime }d\Gamma _{k}^{\prime }=\mu (k)W_{k}  \label{Damp}
\end{equation}%
where
\begin{equation}\label{MU}
\mu (\omega )=2c\int \sigma \left( b,\omega \right) n(r,b)db.
\end{equation}%
Above  $n(r,b)$ is the number density of wormholes at the point $r$ and with the throat radius $b$,
$\sigma \left( b,\omega \right) $ is the total cross-section of such
wormholes (in the most general case it depends on $\omega $), and the multiplier $2$ comes
from taking into account absorption on both throats $S_{+}$ and $S_{-}$
(both give equal contribution due to the symmetry $+\leftrightarrow -$). The
term (\ref{Damp}) simply defines damping along the ray $W_{k}\sim e^{-\tau
}W^0_{k}$, where $\tau =\frac{1}{c}\int \mu d\ell $ is the optical depth.
It is important that in the case of the GW propagation through the normal matter
the damping is determined by the same term (\ref{Damp}) with $\mu (\omega )=c\sigma  _s n_s(r)$,
where $\sigma  _s $ and $n_s$ are the cross-section and the number density of respective objects
(stars, black holes, gas, or any other objects).

The additional radiation capability (secondary sources) is more complex term
which is
\begin{equation}
\delta \alpha _{k}=2c\int \frac{\omega }{\omega ^{\prime }}\delta \left( \xi
_{+}-b\right) \delta \left( \Gamma ^{\prime }-\Gamma _{out}\right) W_k
^{0 \prime } N_{w}
d\Gamma ^{\prime }d\Gamma _{w}.
\end{equation}%
Now, if we use the approximation (due to $\left\vert R_{-}-x^{\prime \prime
}\right\vert \gg \xi _{+}$)%
\begin{equation}
\delta \left( \xi _{+}-b\right) \simeq \pi b^{2}\delta ^{3}\left( \mathbf{x}-%
\mathbf{R}_{+}\right)   \label{delt}
\end{equation}%
which means that in the first approximation throat is a point source at the
position $\mathbf{R}_{+}$, and using the assumption that throat reradiates
in isotropic way (due to averaging over the rotation matrix $U$), then we
find
\begin{equation}
\delta \alpha _{\omega }\simeq \int 
\frac{\omega _{in}}{\omega
_{out}}
\frac{2\pi b^{2}\left[ w\left( \omega _{out},t\right) \right] _{ret}
}{\left\vert \mathbf{R}%
_{-}-\mathbf{x}^{\prime \prime }\right\vert ^{2}}
\delta ^{3}\left(
\mathbf{x}-\mathbf{R}_{+}\right) N_{w}
d\Gamma _{w}.
\end{equation}

Using different distributions for wormholes $N_{w}\left( \Gamma _{w}^{\prime
}\right) $ we may get different answers. The simplest distribution seems to
be (recall that due to the property $\mathbf{V}_{+}=U\mathbf{V}_{-}$ only $%
\mathbf{V}_{+}$ is independent and $V_{+}^{2}=V_{-}^{2}$)
\begin{equation}\label{NW}
N_{w}\left( \Gamma _{w}\right) =\frac{n_{w}}{4\pi \Lambda ^{2}}\delta \left(
\Lambda -\left\vert R_{+}-R_{-}\right\vert \right) \delta \left(
b-b_{0}\right) f(\mathbf{V}_- ),
\end{equation}%
where $ f(\mathbf{V}_{-} ) =( 2\pi \sigma ^2_V) ^{-3/2}\exp ( -
\frac{V_{-} ^{2}}{2\sigma _V ^{2}})$. 
The normalization condition gives $\int N_{w}( \Gamma _{w})
dbd^{3}R_{+}d^{3}R_{-}d^3V_{-}^{3}=n_{w}V$ ($V$ is the volume of space and $n_w$ is the wormhole number 
density).
Such a distribution
corresponds to the case when wormholes are homogeneously distributed in
space and all wormholes have the same throat radius $b=b_{0}$ and the same
distance between entrances $\left\vert R_{+}-R_{-}\right\vert =\Lambda $.

The fact that density $\alpha _{k}$ and, therefore, $W_{k}^{0}$ depend only on $\omega $ simplifies essentially the
re-radiation terms. Let $\sigma _V ^{2}\rightarrow 0$ (i.e., $%
V_{+}^{2}=V_{-}^{2}=0$ and throats are static), then $\omega _{in}=\omega
_{out}$ and we find
\begin{equation}\label{TEQ}
\delta \alpha _{\omega }\simeq \int \frac{2\pi b_{0}^{2}\left[ w\left(
\omega ,t\right) \right] _{ret}}{
\left\vert \mathbf{R}%
_{-}-\mathbf{x}^{\prime \prime }\right\vert ^{2}
}\frac{n_{w}}{4\pi \Lambda ^{2}}\delta \left( \Lambda
-\left\vert \mathbf{x}-\mathbf{R}_{-}\right\vert \right) d^{3}R_{-}.
\end{equation}%
Then for the additional energy density we find 
\begin{equation*}
\delta W_{\omega }\simeq \int \frac{2\pi b_{0}^{2}}{
\left\vert \mathbf{R}%
_{-}-\mathbf{x}^{\prime \prime }\right\vert ^{2}}
\frac{ 
[w( \omega,t)]_{rr}
}{c\left\vert \mathbf{x}-%
\mathbf{R}_{+}\right\vert ^{2}}\delta ^{2}\left( \mathbf{n}_{+}-\mathbf{n}%
_{+}^{\prime }\right) N_{w}dR_{\pm},
\end{equation*}%
where we denote $\mathbf{n}_{+}=\left( \mathbf{x}-%
\mathbf{R}_{+}\right) /\left\vert \mathbf{x}-\mathbf{R}_{+}\right\vert $, 
$\mathbf{n}_{+}^{\prime }=\mathbf{k}/k$, and $dR_{\pm }=d^3R_{+}d^3R_{-}$. 
The function 
$[w( \omega,t)]_{rr}=
w\left( \omega
,t-\frac{\left\vert R_{+}-x\right\vert }{c}-\frac{\left\vert R_{-}-x^{\prime
\prime }\right\vert }{c}\right) 
$ shows that  the
retarding sums. Integrating this over  directions $\mathbf{k}/k=\mathbf{n}%
_{+}^{\prime }$ gives the spectral energy by the relations $\delta
W_{k}d^{3}k=k^{2}\delta W_{k}dkd\Omega ^{\prime }=\frac{\omega ^{2}}{c^{3}}%
\delta W_{\omega }d\omega d\Omega ^{\prime }$ and, therefore, we define the
spectral energy flux  $\delta \widetilde{I}_{\omega }=
\frac{\omega ^{2}}{c^{2}}%
\int \delta W_{\omega }d\Omega ^{\prime }$ as%
\begin{equation}
\begin{array}{ll}
\delta \widetilde{I}_{\omega }\simeq \frac{\omega ^{2}}{c^{2}}\int \frac{%
2\pi b_{0}^{2}}{\left\vert \mathbf{R}_{-}-\mathbf{x}^{\prime \prime
}\right\vert ^{2}}\frac{w\left( \omega ,t-\frac{\left\vert
R_{+}-x\right\vert }{c}-\frac{\left\vert R_{-}-x^{\prime \prime }\right\vert
}{c}\right) }{c\left\vert \mathbf{x}-\mathbf{R}_{+}\right\vert
^{2}}N_{w}dR_{\pm}.
\end{array}
\end{equation}%

In agreement to (\ref{GFE}) for a discrete set of wormholes the  energy flux
can be written as the sum
\begin{equation}\label{echoes:sign}
\begin{array}{ll}
\delta \widetilde{I}_{\omega }\simeq \frac{\omega ^{2}}{c^{2}}\sum \frac{\pi
b_{j}^{2}}{\left\vert \mathbf{R}_{j-}-\mathbf{x}^{\prime \prime }\right\vert
^{2}}\frac{w\left( \omega ,t-\frac{\left\vert R_{j+}-x\right\vert }{c}
-\frac{\left\vert R_{j-}-x^{\prime \prime }\right\vert }{c}\right) }{c\left\vert
\mathbf{x}-\mathbf{R}_{j+}\right\vert ^{2}}+\left(
+\leftrightarrow -\right) ,
\end{array}
\end{equation}
where the sum is taken over wormholes numbered by $j$. Here all cross terms
which may arise from (\ref{GFE}) disappear due to loss of coherence.
Comparing this to the basic energy flux (\ref{signal1}) of the direct signal
\begin{equation}\label{signal2}
I^0_{\omega}(r,t) =\frac{\omega ^{2}}{c^{2}}\frac{w\left( \omega ,t -\frac{\left\vert
\mathbf{r}-\mathbf{r}^{\prime \prime }\right\vert }{c}
\right) }{%
c\left\vert \mathbf{r}-\mathbf{r}^{\prime \prime }\right\vert ^{2}}
\end{equation}%
we see that every echo signal, i.e., every particular term in  (\ref{echoes:sign}),
has the same spectral composition and the same form.
The difference appears only in the arrival times, directions to the source, and amplitudes.

\subsection{Corrections}

The above expressions assume that the space is flat and the red shift is
absent. The redshift can be straightforwardly
accounted for, it gives in (\ref{echoes:sign}) and (\ref{signal2}) the additional multiplier $(1+z)^{-3}$,  
the shift of frequencies, i.e., the replacement
$\omega \rightarrow (1+z)\omega $ in the function $w( \omega)$ (e.g.,  
in the case when $w=w_0\delta (\omega -\omega _0)$
we should replace it with $\frac{w_0}{(1+z)}\delta (\omega -\frac{\omega _0}{1+z})$) and 
the replacement $t\rightarrow a(t)\int dt/a$.
Moreover, they assume also that wormholes have negligible length
of throats, while the substitution (\ref{delt}) produces an error in the
time delay of the order $\eta \sim 2b/c$. In a long throat however the actual flux
(it's amplitude) does not decrease but remains almost constant. From the other
side the spherical symmetry assumes that wormhole metric remains conformally
flat (for the space-like part of the metric). Therefore, formally the
behavior of the energy density scattered by a single wormhole remains the
same $\delta \widetilde{W}_{\omega A}\sim 1/R_{A}^{2}$. All what is actually
changed is the retarding time. This means that the main effect is the adding
an additional retarding time $\eta  _{A}$ to every particular wormhole. From the
rigorous standpoint $\eta _{A}$ can be found by considering geodesics with
the metric (\ref{MT}) and it explicitly depends on all wormhole parameters
and the positions of the source $x''$ and the observer $x$, i.e., $\eta  _{A}=\tau (x,x',\Gamma _{w})$.
Phenomenologically,
however, such a quantity  can be added as an additional parameter to $\Gamma _{w}$, while
(\ref{echoes:sign}) transforms to
\begin{equation}\label{echoes:sign2}
\begin{array}{cc}
\delta \widetilde{I}_{\omega }\simeq \frac{\omega ^{2}}{c^{2}}\sum \frac{\pi
b_{j}^{2}}{\left\vert \mathbf{R}_{j-}-\mathbf{x}^{\prime \prime }\right\vert
^{2}}\frac{w\left( \omega ,t-\frac{\left\vert R_{j+}-x\right\vert }{c}%
-\frac{\left\vert R_{j-}-x^{\prime \prime }\right\vert }{c}-\eta  _{j-}\right) %
}{c\left\vert \mathbf{x}-\mathbf{R}_{j+}\right\vert ^{2}}%
+\\
+\left( +\leftrightarrow -\right) .
\end{array}
\end{equation}

Every particular term in the above expression describes a particular echo signal. On practice,
however, all such signals merge and form a long-living tails for the basic signal. The structure
and the dependence on time of the tail we consider in the next section.

\section{Tails:   time structure of the scattered signal }

Since durations of actual GW signals from binary merges are  very short,
in the leading order every such a signal can be approximated by
delta-like impulse.
Consider the source in the form $w\left( \omega ,t\right)
=w_{0}\left( \omega \right) \delta \left( t-t^{\prime \prime }\right) $. Then the spectral
energy flux $I_{\omega }^{0}( x,t)  $ in
the direct signal is%
\begin{equation}\label{dir:I}
I_{\omega }^{0}\left( x,t\right) =\frac{\omega ^{2}}{c^{3}}\frac{w_{0}\left(
\omega \right) }{R^{2}}\delta \left(t- t^{\prime \prime }-\frac{R}{c}\right)
,
\end{equation}%
where $R^{2}=\left\vert x-x^{\prime \prime }\right\vert ^{2}$, which determines  the values
\begin{equation}\label{Phi0}
\Phi _{\omega }^{0}=
\int I_{\omega }^{0}dt=\frac{\omega ^{2}}{c^{3}}\frac{%
w_{0}\left( \omega \right) }{R^{2}},\ \Phi ^{0}=\int \Phi _{\omega
}^{0}d\omega =\frac{F}{R^{2}},
\end{equation}%
where $4\pi F=4\pi \int \frac{\omega ^{2}}{c^{3}}w_{0}\left( \omega \right)
d\omega $ has the sense of the total energy emitted by the source.

The distribution of wormholes we take in the simplest form (\ref{NW}) but for static wormholes, i.e.,
$\sigma _V\rightarrow 0$ and $N_{w}=\frac{n_{w}%
}{4\pi \Lambda ^{2}}\delta \left( \Lambda -\left\vert \mathbf{R}_{+}-\mathbf{%
R}_{-}\right\vert \right) \delta \left( b-b_{0}\right) $. For simplicity we assume that the delay parameter
$\eta _{\pm}=const$.
Such a
distribution corresponds to the case when all wormholes have the same throat
radius $b=b_{0}$ and the same distance between entrances $\left\vert \mathbf{%
R}_{+}-\mathbf{R}_{-}\right\vert =\Lambda $. The more general case one
obtains by averaging results with an additional distribution $p\left(
\Lambda ,b_{0}\right) $ (which has sense of the probability density for
wormholes $\int p\left( \Lambda ,b_{0}\right) d\Lambda db_{0}=1$).

Let us define the multiplier
\begin{equation}
\beta _{\omega }=\frac{2\pi b_{0}^{2}n_{w}}{4\pi \Lambda ^{2}}R^{2}\Phi
_{\omega }^{0}.
\end{equation}%
Then the additional energy flux $\delta I_{\omega }=c\delta \widetilde{W}%
_{\omega }\ $(spectral energy which falls on a unite square per unit time)
can be cast to the form
\begin{equation}
\begin{array}{ll}
\delta I_{\omega }\simeq \beta _{\omega }\int \frac{\delta \left( t-\frac{%
\left\vert R_{+}-x\right\vert }{c}-\frac{\left\vert R_{-}-x^{\prime \prime
}\right\vert }{c}-\eta  -t^{\prime \prime }\right) 
\delta \left( \Lambda -\left\vert \mathbf{R%
}_{+}-\mathbf{R}_{-}\right\vert \right) 
}{\left\vert
\mathbf{x}-\mathbf{R}_{+}\right\vert ^{2}\left\vert \mathbf{R}_{-}-\mathbf{x}%
^{\prime \prime }\right\vert ^{2}}
dR_{\pm }. 
\end{array}
 \label{FLUX}
\end{equation}%
Fist, we determine the total energy in the tails.

\subsection{Total energy in tails}

The total energy in the tail which falls on a unit square is
\begin{equation*}
\delta \Phi _{\omega }=\int \delta I_{\omega }dt = \beta _{\omega }\int
\frac{\delta \left( \Lambda -\left\vert \mathbf{R}_{+}-\mathbf{R}%
_{-}\right\vert \right) }{\left\vert \mathbf{x}-\mathbf{R}_{+}\right\vert
^{2}\left\vert \mathbf{R}_{-}-\mathbf{x}^{\prime \prime }\right\vert ^{2}}%
dR_{\pm }.
\end{equation*}%
Now using the substitutions  $\mathbf{R}_{+}=\mathbf{X}+\mathbf{R}%
_{-}$ and the set of variables $\mathbf{X}=X\mathbf{n}$,
$\mathbf{R}_{-}=r\mathbf{l}+\mathbf{x}^{\prime
\prime }$, $\mathbf{l}^{2}=\mathbf{n}^2=1$,
we get
\begin{equation*}
\delta \Phi _{\omega }=\frac{\left( 4\pi \right) ^{2}\Lambda ^{2}\beta _{\omega
}}{R}C\left( \chi \right) ,
\end{equation*}%
where  we have defined  variables $R=\left\vert \mathbf{x-x}^{\prime \prime
}\right\vert $,
$\chi =\frac{\Lambda }{R}$, and the function $C( \chi )$ is determined by
the integral \begin{equation}
C\left( \chi \right) \simeq \frac{1}{\left( 4\pi \right) ^{2}}\int \frac{%
d^{2}\Omega _{n}d^{2}\Omega _{l}}{\left\vert \chi \mathbf{n}+y\mathbf{l}-%
\mathbf{m}\right\vert ^{2}}dy.
\end{equation}%
In the above integral we rescale the integration variable $y=\frac{r}{R}$, 
and  $\mathbf{m}=(\mathbf{x}-\mathbf{x}^{\prime \prime })/R$.
This function is determined in Appendix \ref{C(chi)} and  has the form
\begin{equation}
C\left( \chi \right) =
C\left( 0\right)
\left\{
\begin{array}{c}
1,\ \ as\ \chi <1 \\
\frac{1 }{\chi },\ \ as\ \chi >1%
\end{array}%
\right. \allowbreak ,\
\end{equation}
where the numerical constant is $C(0)=2.\,4674$.
This determines  the total energy flux in the tail as
\begin{equation}\label{Etot}
\frac{\delta \Phi _{\omega }}{\Phi _{\omega }^{0}}=4\pi \tau C\left( \chi \right) ,
\end{equation}%
where $\tau =2\pi b_{0}^{2}n_{w}R$. We see that the ratio of amplitudes  is proportional to
$\tau =\tau _w=2\pi \left\langle b^{2}\right\rangle Rn_{w}$, which has sense of the
optical depth, i.e. the mean number of wormholes contained in the volume $%
2\pi b_{0}^{2}R$.  In the case of stars $\tau =\tau _s=\pi \left\langle r_g^{2}\right\rangle Rn_{s}$
and we get $\tau _s\ll 1$
(the multiplier $2$ appears in $\tau_w$ since every wormhole has two entrances).
The mean optical depth determines damping of the basic
signal (\ref{Damp}) when propagating in all possible directions and it
reaches values $\tau _w\gg 1$. It is necessary to stress that the value $\tau _w
\gg 1$ does not mean that the basic signal does not reach the observer,
since in general case wormholes are distributed rather irregularly (or even by a
fractal law). Therefore, there are always directions in which the Universe is
transparent (there are no wormholes on the line of sight). However if we
accept that wormholes are responsible for dark matter phenomenon, then for
 Low Surface Brightness galaxies we find the estimate $e^{\tau _w/2}\sim 10^{3}$, while for mean value
in the Universe it  has at least the order $\tau _w \sim \rho _{DM}/\rho _{b}$
(the ratio of dark matter to baryon densities).

\subsection{The dependence on time}

In the variables $\mathbf{R}_{+}=\mathbf{X}+\mathbf{R}_{-}$ and $\mathbf{R}%
_{-}=\mathbf{R}$ the energy flux (\ref{FLUX}) reads
\begin{equation*}\begin{array}{ll}
\delta I_{\omega }\simeq \beta _{\omega }\int \frac{\delta \left( t-\frac{%
\left\vert X+R-x\right\vert }{c}-\frac{\left\vert
R-x^{\prime \prime }\right\vert }{c}-\eta  -t^{\prime \prime }\right)
\delta \left( \Lambda
-\left\vert \mathbf{X}\right\vert \right) 
}
{\left\vert \mathbf{X}+\mathbf{R}-\mathbf{x}\right\vert ^{2}\left\vert
\mathbf{R}-\mathbf{x}^{\prime \prime }\right\vert ^{2}}
d^{3}\mathbf{X}d^{3}R.
\end{array}
\end{equation*}%
Using coordinates $\mathbf{X=}x\mathbf{n}$, $\mathbf{n}^{2}=1$ and
integrating over $x$ we get
\begin{equation*}
\delta I_{\omega }
\simeq \Lambda ^{2}\beta _{\omega }\int \frac{\delta \left( t-t^{\prime
\prime }-\frac{\left\vert \mathbf{Y+}\Lambda \mathbf{\mathbf{n}}-R\mathbf{m}%
\right\vert }{c}-\frac{Y}{c}-\eta  \right) }{\left\vert \mathbf{Y+}%
\Lambda \mathbf{\mathbf{n}}-R\mathbf{m}\right\vert ^{2}Y^{2}}d^{2}\Omega
_{n}d^{3}Y,
\end{equation*}%
where we use the set of variables as $\mathbf{R}=\mathbf{Y+x}^{\prime \prime }$,
$\mathbf{x}-\mathbf{%
x}^{\prime \prime }=R\mathbf{m}$, and $R=\left\vert \mathbf{x-x}^{\prime \prime
}\right\vert $.
This can be re-written as
\begin{equation}\label{J(xi,chi)}
\delta I_{\omega }\left( x-x^{\prime \prime },t-t^{\prime \prime }\right)
\simeq  c\beta _{\omega }\chi ^2J(\xi ,\chi )
=\Phi _{\omega }^{0}
 \frac{c}{R}
\frac{\tau _w}{4\pi }
J(\xi ,\chi ),
\end{equation}%
where we have denoted $\xi =\frac{\left( t-t^{\prime \prime }-\eta
\right) c}{R}$ , $\chi =\frac{\Lambda }{R}$, and (we use the  variable $\mathbf{y%
}=\mathbf{Y/}R$ in the integral)
\begin{equation}
J(\xi ,\chi )=\int \frac{\delta \left( \left\vert \mathbf{y}-\left( \mathbf{m%
}-\chi \mathbf{\mathbf{n}}\right) \right\vert +y-\xi \right) }{\left\vert
\mathbf{y}-\left( \mathbf{m-}\chi \mathbf{\mathbf{n}}\right) \right\vert ^{2}%
}\frac{d^{3}y}{y^{2}}d^{2}\Omega _{n}.  \label{Ixi}
\end{equation}%
The exact form of $J(\xi ,\chi )$ is determined in the Appendix  \ref{J(0)} and \ref{J}.

\begin{figure}[hbt]
\psfig{figure=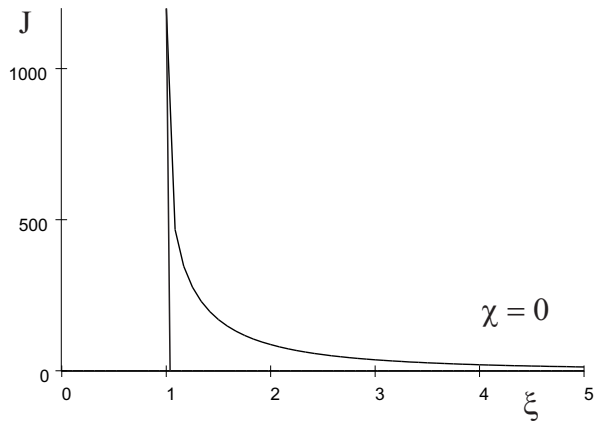,width=7.7cm}
\caption{ The plot of $J(\xi ,0)$ which corresponds to the scattering on normal matter. All the tail lies in the retarded region $\xi >1$. Tails in the case $\chi \ll 1$ have similar form but contain an advanced part.} \label{fig4}
\end{figure}

First, we consider the tail for the scattering by normal matter $\chi =0$.
In the approximation $\chi \ll 1$ and for the region $\xi>1+\chi$ eq. (\ref{Ixi}) gives
\begin{equation}\label{J(xi)}
J(\xi ,\chi )\simeq J(\xi ,0)=8\pi ^{2}\frac{2}{\xi }\ln \frac{\left( \xi +1\right) }{\left\vert \xi
-1\right\vert } .
\end{equation}%
The plot of $J(\xi ,0)$ is given on Fig. \ref{fig4}.
In the limit $\chi
\rightarrow 0$ the distance between throats vanishes and the advanced
signal  is merely absent (e.g., in the region $0<\xi <1$  we get $J(\xi ,0)=0$).
All the scattered signal comes with a
retarding.  The infinite value of $J$ at $\xi =1$ does not mean that the echo is very strong, since it should be compared with the delta impulse
(\ref{dir:I}). In considering  the incident signal of a finite duration $\delta t$, the tail smoothes and takes a finite value.
Indeed, for very small times $\xi -1\ll 1$ we have an approximation $\xi =1+\Delta \xi $ and therefore
$J(\xi ,0)=16\pi ^2\left( \ln 2-\ln \Delta \xi +\frac{\Delta \xi }{2}\ln \frac{%
e\Delta \xi ^{2}}{4}+...\right) $.
Integrating this over $t''$ ($dt''=\Delta t d\xi$, where $\Delta t=R/c$) gives already the finite value
\begin{equation}\label{max}
<J(\xi =1,0)>=\frac{1}{\delta t} \int ^{\delta t}_0J(\xi ,0)dt''\simeq 16\pi ^2  (1+\ln\frac{2\Delta t}{\delta t}).
\end{equation}
Qualitatively, the smoothed function $<J(\xi ,0)>$ behaves as $J(\xi ,\chi )$ for $\chi \ll 1$ which we describe below.
The case $\chi =0$ corresponds  to the scattering on normal matter
when re-radiation has an isotropic character. In this case however the
optical depth $\tau _s \ll 1$ and $\delta I_{\omega }/ I_{\omega
}^{0}\rightarrow 0$. We point out that the value $\tau _s $ has here the statistical character. In the case when there is an object, on the way of the ray,
the echo signal may be considerable. This case is described by lensing or scattering on a single object.

\begin{figure}[hbt]
\psfig{figure=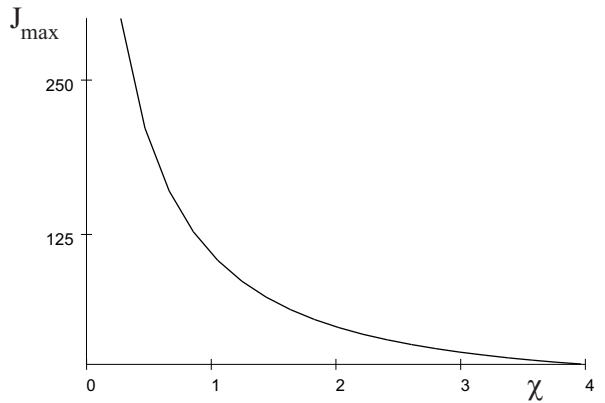,width=7.7cm}
\caption{The plot of $J_{max}(\chi )=J(1+\chi ,\chi )$. This function determines  the maximal possible amplitude in echoes. } \label{fig7}
\end{figure}

\begin{figure}[hbt]
\psfig{figure=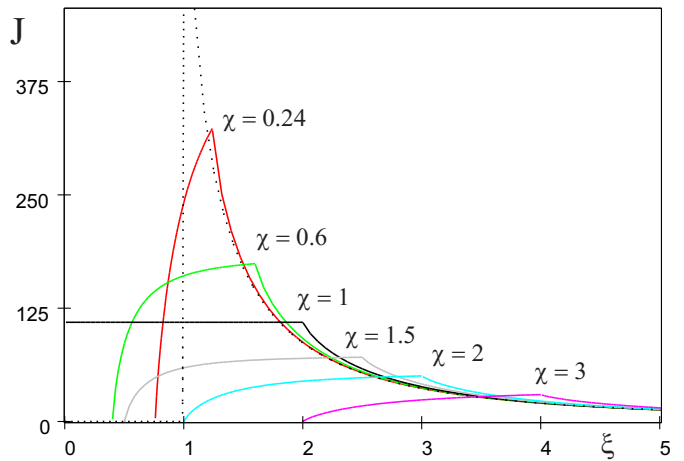,width=8.7cm}
\caption{ The form of tails  $J(\xi ,\chi )$ for a set of values of $\chi $. Advanced part of tails corresponds to $\xi <1$. For values $\chi >2$ tails possess only the retarded part.
} \label{fig5}
\end{figure}

Consider now the case of wormholes. For non-vanishing values $\chi $ the
advanced signal does exist in the region $\xi >1-\chi $ and increases till the maximal value $J_{\max }$ which is reached
 at the point $\xi =1+\chi$. This gives
\begin{equation}
J_{\max }(\chi )=
16\pi ^{2}\left( \frac{\ln \left( \chi
+1\right) }{\chi }-\frac{\ln \chi }{\left( \chi +1\right) }\right) \ .
\end{equation}%
The plot of $J_{max}$ is presented on Fig. \ref{fig7}.
From (\ref{J(xi,chi)}) we see that $J_{max}$ determines the maximal possible amplitude in the echo.
Different forms of tails $J(\xi ,\chi )$ for different values of $\chi  $ are presented on Fig. \ref{fig5}.
Dashed line corresponds to the case $\chi  \rightarrow 0$. The advanced signal corresponds to the region $\xi<1$ and it does exist only for values $\chi <2$. In the region $\chi >2$ the tail is completely retarded.
In the approximation $\chi \gg 1$ we find
\begin{equation}
J(\xi ,\chi )\simeq \frac{1}{\chi ^{2}}J(\frac{\xi }{\chi },0)
=\frac{8\pi ^{2}}{%
\chi }\frac{2 }{\xi }\ln \frac{\left( \xi +\chi \right) }{\left\vert \xi
-\chi \right\vert }\theta \left( \xi -\chi \right) .
\end{equation}%
Here the advanced signal is also absent (since the travel time to wormhole
entrances exceeds the value $\Delta t=c/R$). As we see from Fig. \ref{fig5} for sufficiently big $\xi$ the decrease of tails with time as $\sim 1/\xi \sim 1/t$ is  the common feature of tails for all values of  $\chi $.

\subsection{Structure of general GW  signals with tails}

In the case of a delta-like impulse the total signal is described by
\begin{equation*}
I_{\omega }^{tot}
\left( x,t\right)
=\Phi _{\omega }^{0}\left[ \delta \left( t-t^{\prime \prime
}-\frac{R}{c}\right) +\frac{c}{R}\frac{\tau _{w}}{4\pi }J(\xi ,\chi )\right]
,  \label{I_tot}
\end{equation*}%
where $\Phi _{\omega }^{0}$ is determined by
(\ref{Phi0}).
An arbitrary GW emission we obtain, if we replace $w_0(\omega )$ with a function $w_0( \omega ,t^{\prime \prime })$
and integrate the above equation over $t^{\prime \prime }$.
This defines the spectral energy flux as
\begin{equation*}
I_{\omega }^{tot}
=\frac{\omega ^{2}%
}{c^{3}}\frac{w_{0}\left( \omega ,t-\frac{R}{c}\right) }{R^{2}}
+
\int \frac{\omega ^{2}}{c^{3}}\frac{w_{0}\left( \omega
,t^{\prime \prime }\right) }{R^{2}}\frac{c}{R}\frac{\tau _{w}}{4\pi }J(\xi
,\chi )dt^{\prime \prime }.
\end{equation*}%
The typical duration of emission is very short, while the function $J(\xi ,\chi )$  is very slow function (recall that $\xi =\frac{\left(
t-t^{\prime \prime }-\eta \right) c}{R}$ almost does not change with a small
change of $t^{\prime \prime }$, i.e., $\Delta\xi =c\Delta t^{\prime \prime }/R\ll 1$). This means that the multiplier
$\frac{\tau _{w}}{4\pi }\frac{c}{R}J(\xi
,\chi )$ can be taken at the moment of the start of emission $t^{\prime \prime }=t_0$ and can be removed from the integration.
As it was pointed out above in the case  $\chi =0$ one should replace $J$ with $<J>$  from (\ref{max}) and for simplicity we set $\eta =0$.
Then integrating this over frequencies $I=\int I_{\omega }d\omega $ we find
\begin{equation}\label{DI}
I(t)=I^0(t)+\frac{\tau _{w}}{4\pi }\frac{c}{R}J(\xi _0,\chi )\int
I^{0}(t)dt.
\end{equation}
We see that tails have indeed a universal structure.
 For a short impulse $\int I^0 dt\sim
I^{0}\delta t\sim 2\pi I^{0}/\omega _{0}$, where $\delta t$ is the duration of the basic signal.
The value $R/c=\Delta t$ is the propagation time, and therefore we find
\begin{equation}
\label{ddI}
\frac{\delta I}{I^0}
\simeq \frac{\delta t}{\Delta t}\frac{\tau _{w}}{4\pi }J(\xi _0,\chi )%
\ll 1 .
\end{equation}
The typical ratio $\frac{\delta t}{\Delta t}$ is extremely small $\frac{\delta t}{\Delta t}\sim 10^{-15}$ and, 
therefore, the direct detection of the tail signal is hardly possible.
Indeed, the energy flux relates to the amplitude $h$ as follows
\begin{equation*}
I=\frac{c^{3}}{32\pi G}\frac{1}{T}\int_{T}(\dot{h}_{\times }^{2}+\dot{h}%
_{+}^{2})dt,
\end{equation*}%
where $h_{\times }$ and $h_{+}$ correspond to the two independent polarizations. Assuming the periodic signal $\sim e^{-i\omega _0 t}$ we find
\begin{equation*}
h^{2}+\delta h^2=\frac{16\pi G}{c^{3}\omega _{0}^{2}} I^{0}(t)\left( 1+\frac{\delta I}{I^0}\right) .
\end{equation*}
This explicitly shows that the tail contribution is $\delta h \ll h$.
This means that the echo signals reported in  \cite{echoes1,echoes2,echoes3,echoes4} 
cannot be explained as echoes from wormholes or ordinary compact objects.
However, while the duration of the direct signal $h$ is very short, the tail part $\delta h ^2$ decays with time very slowly $J(\xi _0,\chi )\sim \Delta t/(\Delta t+t)\sim 1$ and the total energy in the tail, according to (\ref{Etot}), may exceed the energy in the basic signal (i.e.,
$\int \delta h ^2 dt\gg \int h^2 dt$). This means that the presence of such a heavy tail may be nevertheless observed by
the increase of the noise level. The tails live for an extremely long period of time, at least $J(\xi _0,\chi )\geq 1$ for $t\sim\Delta t$, and during this time all such tails from different binary merges accumulate. This roughly gives the additional multiplier  $N\simeq \nu\Delta t$, where $\nu$ is the mean rate of events
and this gives already the factor $\delta I/I^0\sim N\delta t\tau _{w}/(4\pi\Delta t ) J(\xi _0,\chi )\gtrsim  10^{-3}$
instead of (\ref{ddI}), where we used for estimates $\nu \delta t \sim 10^{-5} $, $\chi =0.24$, and $\tau _w \sim 5$. Therefore the total noise level can be considerable and the reported in \cite{echoes1,echoes2,echoes3,echoes4} echoes may simply detect the stochastic background of GW radiation.

In conclusion we point out that we have assumed  the specific model of the wormholes distribution when the
distance between wormhole entrances has the same value $\Lambda =\left\vert \mathbf{
R}_{+}-\mathbf{R}_{-}\right\vert $ and all wormholes have the same throat value $b=b_0$. To obtain the more general case we have to
carry out  an additional averaging with some probability density $p\left(
\Lambda ,b_{0}\right) $ which surely somewhat changes the structure of tails. This can be done directly in (\ref{DI}), (\ref{ddI}).


\section{Outlook}
The topological bias for a distribution of wormholes can therefore be evaluated for  Misner-Thorne wormhole by considering the neighborhood of each wormhole in the distribution  after the modelizaition of the Dirac distribution for point-like sources after the characteristic sizes of the throats of each wormhole in the distribution, and by considering the related approximations for the $\beta$ function. Under this modellization, the emission of particles has to be expected within the same range of emission which can be considered not much far form the range of absorption.

The shift on the frequencies is estimated not higher than $10^{-3}$, so that it can be neglected.
Such an estimate is very simple. First, peculiar velocities in galaxies (at edges) are around 300km/s, which gives $v/c \sim 10^{-3}$. This estimate works both for spirals and ellipticals. Second, there are peculiar velocities of clusters (observed by kinematic Sunyaev-Zel'dovich effect) which do not exceed $1500km/s$ which gives the Doppler shift $v/c \sim 5*10^{-3}$.

Rotation curves of spiral galaxies can be observed through several properties, such as Hubble type, structure, activity, and environment \cite{Sofue:2000jx}.
A review on the literature about galaxy rotation curves can be found in \cite{Yegorova:2011fj}, while the possible observational errors are reviewed in \cite{Garcia:2013vma}.
In \cite{Diaferio:2004gk}, the measurements of velocities of clusters due to kinematic Sunyaev-Zel'dovich effect 
are evaluated, for which the systematic errors are examined in \cite{Bhattacharya:2008qc}.

Thus the Doppler shift \cite{deSousa:2002aa} expected does not exceed $0.5\%$, i.e. also at low frequencies \cite{Armstrong:1996rd}. The experimental verification can be modelized at different astrophysical scales, \cite{AJW}, i.e. also at satellite-distances scales \cite{Gerberding:2015xca}.
In the  analysis performed, throats have been assumed to participate in the peculiar motions.

For the local Universe ($z<0.1$) the shift due to the expansion can be estimated to be rather small as well (the frequency shift is less than $10\%$). For the more deep sources the dependence on the redshift can be easily incorporated, e.g., see discussions in Sec. VC. The most distant source GW170729 gives the shift of frequencies which does not exceed $30\%$. 
The behavior of a gravitational wave in several models of expanding universes was studied in \cite{Fabris:1998ak}, while the behavior of a gravitational wave in an expanding Universe in the presence of a point-mass was schematized in \cite{Antoniou:2016vxg}.

Higher-order effects arising from Special Relativity and GR \cite{w93,mag07,w2007} are discussed in \cite{Cooney:2012uc} for comparison with constraints with more general metric theories of gravity and in \cite{galpot,Faber:2005xc}
for comparison with constraints on dark-matter like contributions. We expect that higher-order corrections also do not essentially change the results presented.

In this manner we see that the Doppler shift of frequencies due to particular motions of wormholes can be neglected and the noise can be  estimated to appear in the same frequency range as the original ingoing signal. The dependence on the redshift can not be so easily neglected or eliminated. The explicit form of the tails $J(\xi ,\chi )$ corresponds to the local Universe when the redshift is neglected and for a particular distribution of wormholes having the same distance between entrances into every wormhole $\chi $. In a  more general case we have to make some averaging over possible values of $\chi$.
When the characteristic distance $\chi $ becomes very big the redshift becomes important. In particular, the most distant galaxy  GN-z11 is observed at $z=11.09$ \cite{GN-z11} and we may expect that signals from binary merges may come even from more remote regions with $z>11$.
In other words, together with the noise from the local Universe there should come a  shifted noise from extremely remote regions. This problem however requires the further investigation.

\section{Acknowledgements}

We acknowledge valuable comments and the advice of referees which helped us to clarify some important points and essentially improve the presentation of this work.

\appendix

\section{Function $C( \chi )$}\label{C(chi)}

The function $C( \chi )$ is determined by the integral 
\begin{equation*}
C\left( \chi \right) \simeq \frac{1}{\left( 4\pi \right) ^{2}}\int \frac{%
d^{2}\Omega _{n}d^{2}\Omega _{l}}{\left\vert \chi \mathbf{n}+y\mathbf{l}-%
\mathbf{m}\right\vert ^{2}}dy,
\end{equation*}%
where $d^{2}\Omega $ is the solid angle.
Assuming $\chi =\frac{\Lambda }{R}\ll 1$ we get $C( \chi ) \approx C( 0) $%
\begin{equation}
\begin{array}{ll}
C\left( 0\right) =\frac{1}{2}\int _0^{\infty} \int_0^{\pi}\frac{d(\cos \theta ) }{\left( y^{2}-2y\cos
\theta +1\right) }dy=\\
=\frac{1}{2}\int_{0}^{\infty }\frac{1}{2y}\ln \frac{%
\left( y+1\right) ^{2}}{\left( y-1\right) ^{2}}dy=2.\,\allowbreak 4674 .
\end{array}
\end{equation}%
The opposite case $\chi \gg 1$ reduces to the analogous expression
\begin{equation}
C\left( \chi \right) \simeq \frac{1}{\chi }C\left( 0\right) .
\end{equation}%
In the general case we define  $\mathbf{m}-\chi \mathbf{n}=\gamma \mathbf{k}$, where
$\mathbf{k}^{2}=1$, $
\gamma =\sqrt{\chi ^{2}-2\chi \cos \alpha +1}$,  and $\cos \alpha =(\mathbf{mn})$, then we get
\begin{equation*}
C\left( \chi \right) =\frac{1}{\left( 4\pi \right) ^{2}}\int \left(
\int \frac{d^{2}l}{\left\vert r\mathbf{l}-\mathbf{k}%
\right\vert ^{2}}dr\right) \frac{d^{2}\Omega _{n}}{\gamma },
\end{equation*}%
where $r=\frac{y}{\gamma }$. This transforms the integral into
\begin{equation*}
\begin{array}{ll}
C\left( \chi \right) \simeq \frac{1}{\left( 4\pi \right) ^{2}}\int \left(
\frac{2\pi d(\cos \theta )}{\left( r^{2}-2r\cos \theta +1\right) }dr\right)
\frac{d^{2}\Omega _{n}}{\gamma }=\frac{C\left( 0\right) }{\left( 4\pi
\right) }\int \frac{d^{2}\Omega _{n}}{\gamma }
\end{array}
\end{equation*}%
which gives
\begin{equation*}
C\left( \chi \right) =\frac{C\left( 0\right) }{2}\int_{-1}^{1}\frac{d(\cos
\alpha )}{\sqrt{\chi ^{2}-2\chi \cos \alpha +1}}
.
\end{equation*}%
Finally we find the expression
\begin{equation}\begin{array}{ll}
C\left( \chi \right) =\frac{C\left( 0\right) }{\left( 2\chi \right) }%
 \left( \left( \chi +1\right) -\left\vert \chi -1\right\vert
\right) 
=\frac{C\left( 0\right) }{\chi }\left\{
\begin{array}{c}
\chi ,\ \ as\ \chi <1 \\
1,\ \ as\ \chi >1%
\end{array}%
\right. .
\end{array}
\end{equation}

\section{Function  $J(\xi ,0)$}\label{J(0)}

In the approximation $\chi \rightarrow 0$ eq. (\ref{Ixi}) reduces to%
\begin{equation}
J(\xi ,0)=\int \frac{\delta \left( \left\vert \mathbf{y}-\mathbf{m}%
\right\vert +y-\xi \right) }{\left\vert \mathbf{y}-\mathbf{m}\right\vert ^{2}%
}\frac{d^{3}y}{y^{2}}d^{2}\Omega _{n}.
\end{equation}%
Let us use $x=\cos \theta $, then the above integral gives%
\begin{equation}
J(\xi ,0)=8\pi ^{2}\int_{-1}^{1}\int_{0}^{\infty }\frac{\delta \left( y+%
\sqrt{y^{2}+2yx+1}-\xi \right) }{\left( y^{2}+2yx+1\right) }dydx.
\end{equation}
Let us use new variable
$u=\sqrt{y^{2}+2yx+1}>0$,
$dx=\frac{u}{y}du$ whose
range is
$\left\vert y-1\right\vert <u<y+1$.
Then we get
\begin{equation}
J(\xi ,0)=8\pi ^{2}\int_{0}^{\infty }\left( \int_{\left\vert y-1\right\vert
}^{y+1}\delta \left( y+u-\xi \right) \frac{du}{u}\right) \frac{dy}{y}.
\end{equation}%
This integral splits in two parts which should be considered separately
$J(\xi ,0)=8\pi ^{2}(J_1+J_2)$
\begin{equation*}
\begin{array}{ll}
J_{1}=\int_{0}^{1}\left(
\int_{1-y}^{1+y}\delta \left( y+u-\xi \right) \frac{du}{u}\right) \frac{dy}{y%
} \\
J_{2}=
\int_{1}^{\infty }\left( \int_{y-1}^{y+1}\delta \left( y+u-\xi \right)
\frac{du}{u}\right) \frac{dy}{y}.
\end{array}
\end{equation*}%
Consider the first part $J_{1}$.
 In the regions $\xi <1$ and $ \xi >3$ the delta function has no roots
and we find simply
$J_{1}=0$.
In the rest region $1<\xi <3$  the delta function possesses roots only for
 $y>\frac{1}{2}\left( \xi -1\right)$
 and therefore
\begin{equation*}
J_{1}
=\int_{\frac{1}{2}\left( \xi -1\right) }^{1}\left(
\int_{1-y}^{1+y}\delta \left( u-\left( \xi -y\right) \right) \frac{du}{u}%
\right) \frac{dy}{y},
\end{equation*}%
which defines %
\begin{equation}
J_{1}\left( \xi \right) =\int_{\frac{1}{2}\left( \xi -1\right) }^{1}\frac{dy%
}{\left( \xi -y\right) y}=\left\{
\begin{array}{c}
\allowbreak \frac{1}{\xi }\ln \frac{\left( \xi +1\right) }{\left( \xi
-1\right) ^{2}}\ as\ 1<\xi <3 \\
0\ \ as\ \xi \not\in \lbrack 1,3]%
\end{array}%
\right. .
\end{equation}

Now let us consider the second part $J_2$.
Again in the region $\xi <1$ the delta function has no roots and  $J_{2}=0$.
In the region  $1<\xi <3$ roots exist  for $y<\frac{1%
}{2}\left( \xi +1\right) $ and we find
\begin{equation}
J_{2}
=\int_{1}^{\frac{1}{2}\left( \xi +1\right) }\frac{dy}{\left( \xi
-y\right) y}
=\frac{1}{\xi }\ln \left( \xi +1\right) ,\ as,\
1<\xi <3 \ .
\end{equation}%
In the region  $\xi >3$ we find the upper and lower limits for $y$ as $y=\frac{1}{2}%
\left( \xi \pm 1\right) $ which gives
\begin{equation}
J_{2}=
\int_{\frac{1}{2}\left( \xi -1\right) }^{\frac{1}{2}\left( \xi
+1\right) }\frac{dy}{\left( \xi -y\right) y}
=\frac{1}{\xi }\ln \frac{\left( \xi +1\right) ^{2}}{%
\left( \xi -1\right) ^{2}},\ as,\ \xi >3 \ .
\end{equation}%
Now collecting all we find
\begin{equation}
J(\xi ,0)=8\pi ^{2}\frac{2}{\xi }\ln \frac{\left( \xi +1\right) }{\left( \xi
-1\right) }\theta (\xi -1),  \label{I(0)}
\end{equation}%
where $\theta ( \xi -1)  $ is the step function.
We also point out that $\int J(\xi ,0)d\xi
=
(4\pi )^2 C(0)=8\pi ^2\times 4.\,\allowbreak 9348$.

\section{Function $J(\xi ,\chi )$}\label{J}

Consider the  function $J(\xi ,\chi )$ which is determined by the
integral (\ref{Ixi}).
\begin{equation*}
J(\xi ,\chi )=\int \frac{\delta \left( \left\vert \mathbf{y}-\left( \mathbf{%
m-}\chi \mathbf{\mathbf{n}}\right) \right\vert +y-\xi \right) }{\left\vert
\mathbf{y}-\left( \mathbf{m-}\chi \mathbf{\mathbf{n}}\right) \right\vert ^{2}%
}\frac{d^{3}y}{y^{2}}d^{2}\Omega _{n}  .
\end{equation*}%
If we define new variables as $\mathbf{m}-\chi \mathbf{n}=\gamma \mathbf{k}$
with $\gamma =\sqrt{\chi ^{2}-2\chi \cos \alpha +1}$,  $\cos \alpha =(\mathbf{mn})$, and $r=y/\gamma $,
then the integral  reduces to the function $J(\xi ,0)$ as
\begin{equation*}
   \begin{array}{ll} 
J(\xi ,\chi )=\int \left( \int \frac{\delta \left( \left\vert \mathbf{r}-%
\mathbf{k}\right\vert +r-\frac{\xi }{\gamma }\right) }{\left\vert \mathbf{r}-\mathbf{k%
}\right\vert ^{2}}\frac{d^{3}r}{r^{2}}\right) \frac{d^{2}\Omega _{n}}{\gamma
^{2}} 
= 
\\
=\frac{1}{2}\int_{-1}^{1}J(\frac{\xi }{\gamma },0)\frac{d(\cos \alpha )}{%
\gamma ^{2}}.  
 \end{array}
\end{equation*}%
Now by means of introducing $\gamma $ as a new integration variable instead
of $\cos \alpha $ we find%
\begin{equation}
J(\xi ,\chi )=\frac{1}{2\chi }\int_{\left\vert \chi -1\right\vert }^{\chi
+1}J(\frac{\xi }{\gamma },0)\frac{d\gamma }{\gamma }.  \label{Itot}
\end{equation}
Substituting here (\ref{I(0)}) we get
\begin{equation}
J(\xi ,\chi )
=\frac{8\pi ^{2}}{\chi }\int_{\frac{\left\vert \chi -1\right\vert }{%
\xi }}^{\frac{\chi +1}{\xi }}\ln \frac{\left( 1+x\right) }{\left( 1-x\right)
}\theta \left( 1-x\right)dx,
\end{equation}%
where $\theta \left( 1 -x \right) $ is the step function.
This integral has different forms for different regions.

\emph{First region } is $\xi <\left\vert \chi -1\right\vert $. It  gives  $x>1
$ and, therefore,
\begin{equation}
J(\xi ,\chi )=0,\ \ as\ \ \xi <\left\vert \chi -1\right\vert
\end{equation}

\emph{The second region } is $\left\vert \chi -1\right\vert <\xi <\chi +1$. The step function defines the upper limit $x=1$ and
we get
\begin{equation}
\begin{array}{ll}
J(\xi ,\chi )=\frac{8\pi ^{2}}{\chi }
\left(  \ln \left(
\frac{4\xi ^{2}}{\xi ^{2}-\left( \chi -1\right) ^{2}}\right) -\frac{%
\left\vert \chi -1\right\vert }{\xi }\ln \frac{\left( \xi +\left\vert \chi
-1\right\vert \right) }{\left( \xi -\left\vert \chi -1\right\vert \right) }%
\right).
\end{array}
\end{equation}%

\emph{The last region } is $\xi >\chi +1$. Then $\theta \left( 1 -x \right) =1$ and we find
\begin{equation}
\begin{array}{ll}
J
=\frac{8\pi ^{2}}{\chi }
\left( \ln \frac{\xi ^{2}-\left( \chi
+1\right) ^{2}}{\xi ^{2}-\left( \chi -1\right) ^{2}}+\frac{\chi +1}{\xi }\ln
\frac{\left( \xi +\chi +1\right) }{\left( \xi -\chi -1\right) }-\frac{%
\left\vert \chi -1\right\vert }{\xi }\ln \frac{\left( \xi +\left\vert \chi
-1\right\vert \right) }{\left( \xi -\left\vert \chi -1\right\vert \right) }%
\right) .
\end{array}
\end{equation}

\section{Free motion}\label{freem}

For a point-like source $\alpha _{k}=w_{0}\left( \omega \right) \delta
\left( x-x^{\prime \prime }\right) \delta \left( t-t^{\prime \prime }\right)
$ and in the absence of the scattering on wormholes (i.e., in the case of
free motion) eq. (\ref{FEE}) can be solved by the characteristics method.
Indeed the geodesic motion of particles is described by rays
\begin{equation}
\mathbf{r}=\mathbf{r}_{0}+\mathbf{V}_{g}(t-t^{\prime \prime }),\ \mathbf{k}=%
\mathbf{k}_{0},\text{ }\mathbf{V}_{g}=\frac{\partial \omega }{\partial
\mathbf{k}}=c\frac{\mathbf{k}}{k}.
\end{equation}%
Then, if we take $(k_{0},r_{0})$ as new coordinates, the energy density
becomes $W\left[ k\left( k_{0},r_{0},t\right) ,r\left( k_{0},r_{0},t\right) %
\right] =W\left( k_{0},r_{0},t\right) $ and (\ref{FEE}) reads
\begin{equation}
\frac{dW\left( k_{0},r_{0},t\right) }{dt}=\alpha _{k}=w_{0}\left( \omega
\right) \delta \left( \mathbf{r}_{0}-\mathbf{r}^{\prime \prime }\right)
\delta \left( t-t^{\prime \prime }\right) ,
\end{equation}%
which has the obvious solution in the form
\begin{equation}
W\left( k_{0},r_{0},t\right) =w_{0}\left( \omega \right) \delta \left(
\mathbf{r}_{0}-\mathbf{r}^{\prime \prime }\right) \theta \left( t-t^{\prime
\prime }\right) ,
\end{equation}%
where $\theta \left( t-t^{\prime \prime }\right) $ is the step function ( $%
\theta \left( x\right) =1$ as $x>0$). Now to return to the initial
coordinates $\left( k,r\right) $ we should simply replace back
\begin{equation}
\mathbf{r}_{0}=\mathbf{r}-\mathbf{V}_{g}(t-t^{\prime \prime }),\ \ \mathbf{k}%
_{0}=\mathbf{k},
\end{equation}%
which gives (we assume $t>t^{\prime \prime }$)%
\begin{equation}
W_k\left( r,t\right) =w_{0}\left( \omega \right) \delta \left( \mathbf{r}-%
\mathbf{V}_{g}(t-t^{\prime \prime })-\mathbf{r}^{\prime \prime }\right) .
\end{equation}%
In terms of the spherical coordinates $\mathbf{V}_{g}=(c,\theta ^{\prime
},\phi ^{\prime })$ and $\mathbf{r}-\mathbf{r}^{\prime \prime }=\left(
\left\vert \mathbf{r}-\mathbf{r}^{\prime \prime }\right\vert ,\theta ,\phi
\right) $ the above expression transforms to%
\begin{equation*}
 \begin{array}{cc} 
W_k( r,t) =\frac{w_{0}\left( \omega \right) }{c\left\vert \mathbf{%
r}-\mathbf{r}^{\prime \prime }\right\vert ^{2}}\times 
\\
\times
\delta \left( \frac{%
\left\vert \mathbf{r}-\mathbf{r}^{\prime \prime }\right\vert }{c}%
-t+t^{\prime \prime }\right) \delta \left( \cos \theta -\cos \theta ^{\prime
}\right) \delta \left( \phi -\phi ^{\prime }\right) .
 \end{array}
\end{equation*}

\end{document}